\documentclass[aps,showpacs,superscriptaddress,12pt,tightenlines]{revtex4}
\usepackage{times}
\usepackage{amsmath}
\usepackage{graphicx}
\usepackage{epsfig}
\usepackage{dcolumn}
\usepackage{bm}
\usepackage{color}

\newcommand{\BR}{{\cal B}}

\newcommand{\pp}{\pi^+\pi^-}
\newcommand{\ks}{K_S^0}

\newcommand{\EE}{e^+e^-}

\newcommand{\psp}{\psi(3686)}

\newcommand{\jpsi}{J/\psi}
\newcommand{\piz}{\pi^0}

\newcommand{\zcp}{Z_c(4020)}

\newcommand{\ppjpsi}{\pi^+\pi^-J/\psi}
\newcommand{\hc}{h_c}
\newcommand{\pphc}{\pi^+\pi^-\hc}

\newcommand{\psifty}{\psi(4040)}
\newcommand{\psifto}{\psi(4160)}
\newcommand{\psiftf}{\psi(4415)}

\def\Journal#1#2#3#4{{#1} {\bf #2}, #3 (#4)}

\def\PRL{Phys. Rev. Lett.}
\def\PRD{Phys. Rev. D}

\def\EPJC{Eur. Phys. J. C}

\begin{document}

\title{\boldmath
Evidence for resonant structures in $\EE\to \pp\hc$}

\author{Chang-Zheng Yuan}
\affiliation{Institute of High Energy Physics,\\
Chinese Academy of Sciences, Beijing 100049, China}

\date{\today}

\begin{abstract}

The cross sections of $\EE\to \pphc$ at center-of-mass energies
from 3.90 to 4.42~GeV were measured by the BESIII and the CLEO-c
experiments. Resonant structures are evident in the $\EE\to \pphc$
line shape, the fit to the line shape results in a narrow
structure at a mass of $(4216\pm 18)~{\rm MeV}/c^2$ and a width of
$(39\pm 32)$~MeV, and a possible wide structure of mass $(4293\pm
9)~{\rm MeV}/c^2$ and width $(222\pm 67)$~MeV. Here the errors are
combined statistical and systematic errors. This may indicate that
the $Y(4260)$ state observed in $\EE\to \ppjpsi$ has fine
structure in it.

\end{abstract}


\maketitle

The observation of the $Y$-states in the exclusive production of
$\pp\jpsi$~\cite{babary,belley,babary_new,belley_new} and
$\pp\psp$~\cite{babar_pppsp,belle_pppsp,babar_pppsp_new} from the
B-factories is a great puzzle in understanding the vector
charmonium states~\cite{epjc-review}. According to the potential
models, there are 5 vector states above the well-known 1D state
$\psi(3770)$ and below around 4.7~GeV/$c^2$, namely, the 3S, 2D,
4S, 3D, and 5S states~\cite{epjc-review}. However, experimentally,
besides the three well known structures observed in inclusive
hadronic cross section, i.e., the $\psifty$, $\psifto$, and
$\psiftf$~\cite{pdg}, there are four $Y$-states, i.e., the
$Y(4008)$, $Y(4260)$, $Y(4360)$, and
$Y(4660)$~\cite{babary,belley,babary_new,belley_new, babar_pppsp,
belle_pppsp,babar_pppsp_new}. This suggests that at least some of
these structures are not charmonium states, and thus has arisen
various scenarios in interpreting one or more of
them~\cite{epjc-review}.

The BESIII experiment~\cite{bes3} running near the open charm
threshold supplies further information to understand the
properties of these vector states. Amongst these information, the
most relevant measurement is the study of $\EE\to \pp
h_c$~\cite{zc4020}. Besides the observation of a charged
charmoniumlike state $\zcp$, BESIII reported the cross section
measurement of $\EE\to \pp h_c$ at 13 center-of-mass (CM) energies
from 3.900 to 4.420~GeV~\cite{zc4020}. The measurements are listed
in Table~\ref{scan-data}. In the studies, the $h_c$ is
reconstructed via its electric-dipole (E1) transition $h_c\to
\gamma\eta_c$ with $\eta_c$ to 16 exclusive hadronic final states:
$p \bar{p}$, $2(\pi^+ \pi^-)$, $2(K^+ K^-)$, $K^+ K^- \pi^+
\pi^-$, $p \bar{p} \pi^+ \pi^-$, $3(\pi^+ \pi^-)$, $K^+ K^-
2(\pi^+ \pi^-)$, $\ks K^\pm \pi^\mp$, $\ks K^\pm \pi^\mp \pi^\pm
\pi^\mp$, $K^+ K^- \pi^0$, $p \bar{p}\pi^0$, $\pi^+ \pi^- \eta$,
$K^+ K^- \eta$, $2(\pi^+ \pi^-) \eta$, $\pi^+ \pi^- \pi^0 \pi^0$,
and $2(\pi^+ \pi^-) \pi^0 \pi^0$.

\begin{table}[htbp]
\caption{$\EE\to \pphc$ cross sections measured from the BESIII
experiment. For the first three energy points, besides the upper
limits, the central values and the statistical errors which will
be used in the fits below are also listed. The second errors are
systematic errors and the third ones are from the uncertainty in
$\BR(h_c\to \gamma\eta_c)$~\cite{bes3-hc-inclusive}.}
\label{scan-data} \centering
\begin{tabular}{cc}
  \hline\hline
  $\sqrt{s}$~(GeV) &  $\sigma(\EE\to \pphc)$~(pb) \\
  \hline
  3.900    & $0.0\pm 6.0$ or $<8.3$ \\
  4.009    & $1.9\pm 1.9$ or $<5.0$ \\
  4.090    & $0.0\pm 7.4$ or $<13$ \\
  4.190    & $17.7\pm  9.8\pm  1.6\pm 2.8$ \\
  4.210    & $34.8\pm  9.5\pm  3.2\pm 5.5$ \\
  4.220    & $41.9\pm 10.7\pm  3.8\pm 6.6$ \\
  4.230    & $50.2\pm  2.7\pm  4.6\pm 7.9$ \\
  4.245    & $32.7\pm 10.3\pm  3.0\pm 5.1$ \\
  4.260    & $41.0\pm  2.8\pm  3.7\pm 6.4$ \\
  4.310    & $61.9\pm 12.9\pm  5.6\pm 9.7$ \\
  4.360    & $52.3\pm  3.7\pm  4.8\pm 8.2$ \\
  4.390    & $41.8\pm 10.8\pm  3.8\pm 6.6$ \\
  4.420    & $49.4\pm 12.4\pm  4.5\pm 7.6$ \\
  \hline\hline
\end{tabular}
\end{table}

The CLEO-c experiment did a similar analysis, but with significant
signal only at CM energy 4.17~GeV~\cite{cleoc_pipihc}, the result
is $\sigma=(15.6\pm 2.3\pm 1.9\pm 3.0)$~pb, where the third error
is from the uncertainty in $\BR[\psp\to \piz\hc]$.

\begin{figure}[bp]
\centering
 \includegraphics[width=10cm]{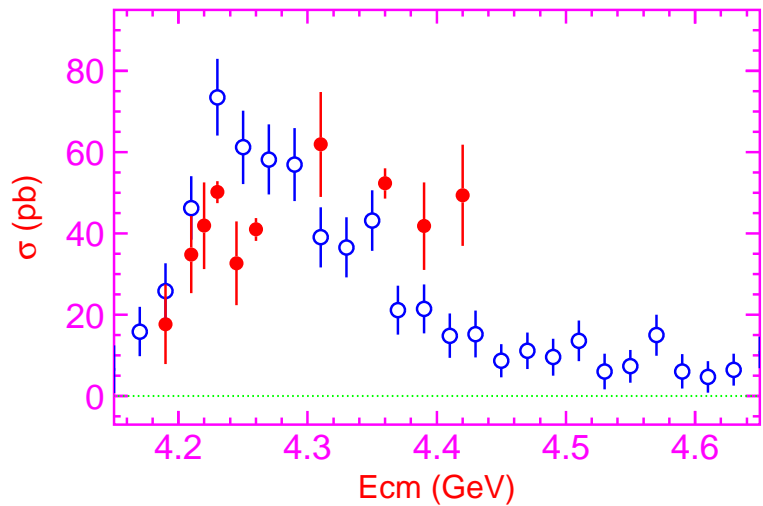}
\caption{The comparison between the cross sections of $\EE\to \pp
h_c$ from BESIII~(dots with error bars)~\cite{zc4020} and those of
$\EE\to \ppjpsi$ from Belle~(open circles with error
bars)~\cite{belley_new}. The errors are statistical only.}
  \label{pphc_ppjpsi}
\end{figure}

The cross sections are of the same order of magnitude as those of
the $\EE\to \ppjpsi$ measured by BESIII~\cite{zc3900} and other
experiments~\cite{babary_new,belley_new}, but with a different
line shape (see Fig.~\ref{pphc_ppjpsi}). There is a broad
structure at high energy with a possible local maximum at around
4.23~GeV. We try to use the BESIII and the CLEO-c measurements to
extract the resonant structures in $\EE\to \pp\hc$.

As the systematic error ($\pm 18.1\%$) of the BESIII experiment is
common for all the data points, we only use the statistical errors
in the fits below. The CLEO-c measurement is completely
independent from the BESIII experiment, and all the errors added
in quadrature ($\pm 4.2$~pb) is taken as the total error and is
used in the fits. We use a least $\chi^2$ method
with~\cite{footnote}
$$
\chi^2 = \sum_{i=1}^{14}
         \frac{(\sigma^{\rm meas}_i-\sigma^{\rm fit}(m_i))^2}
              {(\Delta\sigma^{\rm meas}_i)^2},
$$
where $\sigma^{\rm meas}_i\pm \Delta\sigma^{\rm meas}_i$ is the
experimental measurement, and $\sigma^{\rm fit}(m_i)$ is the cross
section value calculated from the model below with the parameters
from the fit. Here $m_i$ is the energy corresponds to the $i$th
energy point.

As the line shape above 4.42~GeV is unknown, it is not clear
whether the large cross section at high energy will decrease or
not. We try to fit the data with two different scenarios.

Assuming the cross section follows the three-body phase space and
there is a narrow resonance at around 4.2~GeV, we fit the cross
sections with the coherent sum of two amplitudes, a constant and a
constant width relativistic Breit-Wigner (BW) function, i.e.,
$$
 \sigma(m)=|c\cdot \sqrt{PS(m)} + e^{i\phi}BW(m)\sqrt{PS(m)/PS(M)}|^2,
$$
where $PS(m)$ is the 3-body phase space factor,
$BW(m)=\frac{\sqrt{12\pi\Gamma_{\EE}\BR(\pp\hc)\Gamma_{\rm tot}}}
{m^2-M^2+iM\Gamma_{\rm tot}}$, is the Breit-Wigner (BW) function
for a vector state, with mass $M$, total width $\Gamma_{\rm tot}$,
electron partial width $\Gamma_{\EE}$, and the branching fraction
to $\pp\hc$, $\BR(\pp\hc)$, keep in mind that from the fit we can
only extract the product $\Gamma_{\EE}\BR(\pp\hc)$. The constant
term $c$ and the relative phase, $\phi$, between the two
amplitudes are also free parameters in the fit together with the
resonant parameters of the BW function.

\begin{figure}[htbp]
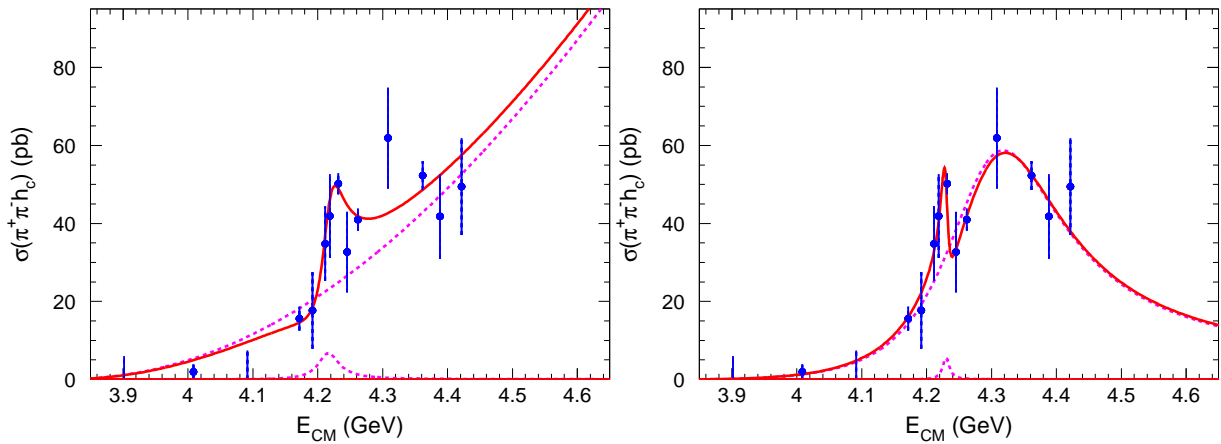

\centering
 \includegraphics[height=8cm,angle=-90]{BW+PS.epsi}
 \includegraphics[height=8cm,angle=-90]{2BW.epsi}
 \caption{The fit to the cross sections of $\EE\to \pp
h_c$ from BESIII and CLEO-c~(dots with error bars). Solid curves
show the best fits, and the dashed ones are individual component.
Left panel is the fit with the coherent sum of a phase space
amplitude and a BW function, and the right panel is the coherent
sum of two BW functions.}
  \label{fit_pipihc}
\end{figure}

The fit indicates the existence of a resonance (called $Y(4220)$
hereafter) with a mass of $(4216\pm 7)$~MeV/$c^2$ and width of
$(39\pm 17)$~MeV, and the goodness-of-the-fit is $\chi^2/{\rm ndf}
= 11.04/9$, corresponding to a confidence level of 27\%. There are
two solutions for the $\Gamma_{\EE}\times {\cal B}(Y(4220)\to
\pphc)$ which are $(0.32\pm 0.15)$~eV and $(6.0\pm 2.4)$~eV. Here
all the errors are from fit only. Fitting the cross sections
without the $Y(4220)$ results in a very bad fit, $\chi^2/{\rm ndf}
= 72.75/13$, corresponding to a confidence level of $2.5\times
10^{-10}$. The statistical significance of the $Y(4220)$ is
calculated to be $7.1\sigma$ comparing the two $\chi^2$s obtained
above and taking into account the change of the
number-of-degree-of-freedom. Figure~\ref{fit_pipihc}~(left panel)
shows the final fit with the $Y(4220)$.

Assuming the cross section decreases at high energy, we fit the
cross sections with the coherent sum of two constant width
relativistic BW functions, i.e.,
$$
 \sigma(m) = |BW_1(m)\cdot \sqrt{PS(m)/PS(M_1)}
   + e^{i\phi}BW_2(m)\cdot \sqrt{PS(m)/PS(M_2)}|^2,
$$
where both $BW_1$ and $BW_2$ take the same form as $BW(m)$ above
but with different resonant parameters.

The fit indicates the existence of the $Y(4220)$ with a mass of
$(4230\pm 10)$~MeV/$c^2$ and width of $(12\pm 36)$~MeV, as well as
a broad resonance, the $Y(4290)$, with a mass of $(4293\pm
9)$~MeV/$c^2$ and width of $(222\pm 67)$~MeV. The
goodness-of-the-fit is $\chi^2/{\rm ndf} = 1.81/7$, corresponding
to a confidence level of 97\%, an almost perfect fit. There are
two solutions for the $\Gamma_{\EE}\times {\cal
B}[Y(4220)/Y(4290)\to \pphc]$ which are $(0.07\pm 0.07)~{\rm
eV}/(16.1\pm 2.2)$~eV and $(2.7\pm 4.9)~{\rm eV}/(19.0\pm
5.9)$~eV. Again, here the errors are from fit only. Fitting the
cross sections without the $Y(4220)$ results in a much worse fit,
$\chi^2/{\rm ndf} = 30.65/11$, corresponding to a confidence level
of $1.3\times 10^{-3}$. The statistical significance of the
$Y(4220)$ is calculated to be $4.5\sigma$ comparing the two
$\chi^2$s obtained above and taking into account the change of the
number-of-degree-of-freedom. Figure~\ref{fit_pipihc}~(right panel)
shows the final fit with the $Y(4220)$ and $Y(4290)$.

From the two fits showed above, we conclude that very likely there
is a narrow structure at around 4.22~GeV/$c^2$, although we are
not sure if there is a broad resonance at 4.29~GeV/$c^2$. We try
to average the results from the fits to give the best estimation
of the resonant parameters. For the $Y(4220)$, we obtain
\begin{eqnarray*}
M(Y(4220))                &=& (4216\pm 18)~\hbox{MeV}/c^2, \\
\Gamma_{\rm tot}(Y(4220)) &=& (39\pm 32)~\hbox{MeV}, \\
\Gamma^{Y(4220)}_{\EE}\times \BR[Y(4220)\to \pphc]
                          &=& (4.6\pm 4.6)~\hbox{eV}.
\end{eqnarray*}
While for the $Y(4290)$, we obtain
\begin{eqnarray*}
M(Y(4290))                &=& (4293\pm 9)~\hbox{MeV}/c^2, \\
\Gamma_{\rm tot}(Y(4290)) &=& (222\pm 67)~\hbox{MeV}, \\
\Gamma^{Y(4290)}_{\EE}\times \BR[Y(4290)\to \pphc]
                          &=& (18\pm 8)~\hbox{eV}.
\end{eqnarray*}
Here the errors include both statistical and systematic errors.
The results from the two solutions and the two fit scenarios are
covered by enlarged errors, the common systematic error in the
cross section measurement is included in the error of the
$\Gamma_{\EE}$.

It is noticed that the uncertainties of the resonant parameters of
the $Y(4220)$ are large, this is due to two important facts: one
is the lack of data at CM energies above 4.42~GeV which may
discriminate which of the two above scenarios is correct, the
other is the lack of high precision measurements around the
$Y(4220)$ peak, especially between 4.23 and 4.26~GeV. The two-fold
ambiguity in the fits is a nature consequence of the coherent sum
of two amplitudes~\cite{zhuk}, although high precision data will
not resolve the problem, they will reduce the errors in
$\Gamma_{\EE}$ from the above fits. As the fit with a phase space
amplitude predicts rapidly increasing cross section at high
energy, it is very unlikely to be true, so the results from the
fit with two resonances is more likely to be true. More
measurements from the BESIII experiments at CM energies above
4.42~GeV and more precise data at around the $Y(4220)$ peak will
also be crucial to settle down all these problems.

There are thresholds of $\bar{D}D_1$~\cite{zhaoq1},
$\omega\chi_{cJ}$~\cite{zhenghq,yuancz},
$D_s^{\ast+}D_s^{\ast-}$~\cite{pdg} at the $Y(4220)$ mass region,
these make the identification of the nature of this structure very
complicated. The fits described in this paper supply only one
possibility of interpreting the data. In Ref.~\cite{zhaoq2}, the
BESIII measurements~\cite{zc4020} were described with the presence
of one relative S-wave $\bar{D} D_1 + c.c.$ molecular state
$Y(4260)$ and a non-resonant background term; while in
Ref.~\cite{voloshin}, the BESIII data~\cite{zc4020} were fitted
with a model where the $Y(4260)$ and $Y(4360)$ are interpreted as
the mixture of two hadroncharmonium states. It is worth to point
out that various QCD calculations indicate that the
charmonium-hybrid lies in the mass region of these two $Y$
states~\cite{ccg_lqcd} and the $c\bar{c}$ tend to be in a
spin-singlet state. Such a state may couple to a spin-singlet
charmonium state such as $h_c$ strongly, this makes the $Y(4220)$
and/or $Y(4290)$ good candidates for the charmonium-hybrid states.

In summary, we fit $\EE\to \pp\hc$ cross sections measured by
BESIII and CLEO-c experiments, evidence for a narrow structure at
around 4.22~GeV, as well as a wide one at 4.29~GeV, is observed.
More high precision measurements at above 4.42~GeV and around
4.22~GeV are desired to better understand these structures.

%%%%%%%%%%%%%%%%%%%%%%%%%%%%%%%%%%%%%
%%%% Acknowledge add later %%%%%%%%%%
%%%%%%%%%%%%%%%%%%%%%%%%%%%%%%%%%%%%%

This work was supported in part by the Ministry of Science and
Technology of China under Contract No. 2009CB825203, and National
Natural Science Foundation of China (NSFC) under Contracts Nos.
10825524, 10935008, and 11235011.

%%%%%%%%%%%%%%%%%%%%%%%%%%%%%%%%%%%%%
%%%%%%%%%%%%%%%%%%%%%%%%%%%%%%%%%%%%%


\begin{thebibliography}{**}

\bibitem{babary} B. Aubert {\em et al.} (BaBar Collaboration).
Phys. Rev. Lett. {\bf 95}, 142001 (2005).

\bibitem{belley} C.~Z. Yuan {\em et al.} (Belle Collaboration).
Phys. Rev. Lett. {\bf 99}, 182004 (2007).

\bibitem{babary_new} J.~P.~Lees {\em et al.} (BaBar Collaboration),
\Journal\PRD{86}{051102(R)}{2012}.

\bibitem{belley_new} Z.~Q.~Liu {\em et al.} (Belle
Collaboration), \Journal\PRL{110}{252002}{2013}.

\bibitem{babar_pppsp} B. Aubert {\em et al.} (BaBar Collaboration).
Phys. Rev. Lett. {\bf 98}, 212001 (2007).

\bibitem{belle_pppsp} X.~L. Wang {\em et al.} (Belle Collaboration).
Phys. Rev. Lett. {\bf 99}, 142002 (2007).

\bibitem{babar_pppsp_new} B. Aubert {\em et al.} (BaBar Collaboration).
arXiv:1211.6271.

\bibitem{epjc-review} For a recent review,
see N. Brambilla {\em et al.}, \Journal\EPJC{71}{1534}{2011}.

\bibitem{pdg} J. Beringer {\em et al.} (Particle Data Group),
\Journal\PRD{86}{010001}{2012}.

\bibitem{bes3} M. Ablikim {\em et al.} (BESIII Collaboration),
Nucl. Instrum. Methods Phys. Res., Sect. A {\bf 614}, 345 (2010).

\bibitem{zc4020} M.~Ablikim {\em et al.} (BESIII Collaboration),
\Journal\PRL{111}{242001}{2013}.

\bibitem{bes3-hc-inclusive} M.~Ablikim {\em et al.} (BESIII
Collobarotion), Phys.\ Rev.\ Lett.\ {\bf 104}, 132002 (2010).

\bibitem{cleoc_pipihc}
  T.~K.~Pedlar {\it et al.}  (CLEO Collaboration),
  %``Observation of the $h_c(1P)$ using $e^+e^-$ collisions
  % above $D\bar{D}$ threshold,''
  Phys.\ Rev.\ Lett.\  {\bf 107}, 041803 (2011).

\bibitem{zc3900} M.~Ablikim {\em et al.} (BESIII Collaboration),
\Journal\PRL{110}{252001}{2013}.

\bibitem{footnote} For the three low statistics energy points,
the $\chi^2$ is not well defined. We take the central values
listed in Table~\ref{scan-data} as nominal values, and vary the
central values and statistical errors in a wide range to estimate
the possible bias in this assumption. The bias is found to be
small and is considered as systematic error of the results.

\bibitem{zhuk}
  K.~Zhu, X.~H.~Mo, C.~Z.~Yuan and P.~Wang,
  %``A mathematical review on the multiple-solution problem,''
  Int.\ J.\ Mod.\ Phys.\ A {\bf 26}, 4511 (2011).

\bibitem{zhaoq1}
Q.~Wang, C.~Hanhart and Q.~Zhao, Phys.\ Rev.\ Lett.\  {\bf 111},
132003 (2013).

\bibitem{zhenghq}
  L.~Y.~Dai, M.~Shi, G.~-Y.~Tang and H.~Q.~Zheng,
  %``On the Nature of X(4260),''
  arXiv:1206.6911 [hep-ph].
  %%CITATION = ARXIV:1206.6911;%%

\bibitem{yuancz}
  C.~Z.~Yuan, P.~Wang and X.~H.~Mo,
  %``The Y(4260) as an omega chi(c1) molecular state,''
  Phys.\ Lett.\ B {\bf 634}, 399 (2006).

\bibitem{zhaoq2}
M.~Cleven, Q.~Wang, F.~-K.~Guo, C.~Hanhart, U.~-G.~Mei\ss ner and
Q.~Zhao,
  %``$\bm{Y(4260)}$ as the first $S$-wave open charm vector
  % molecular state,''
  arXiv:1310.2190 [hep-ph].

\bibitem{voloshin}
  X.~Li and M.~B.~Voloshin,
  %``Y(4260) and Y(4360) as mixed hadrocharmonium,''
  arXiv:1309.1681 [hep-ph].
  %%CITATION = ARXIV:1309.1681;%%

\bibitem{ccg_lqcd}
  T.~Barnes, F.~E.~Close and E.~S.~Swanson,
  %``Hybrid and conventional mesons in the flux tube model:
  % Numerical studies and their phenomenological implications,''
  Phys.\ Rev.\ D {\bf 52}, 5242 (1995);
  P.~Guo, A.~P.~Szczepaniak, G.~Galata, A.~Vassallo and E.~Santopinto,
  %``Heavy quarkonium hybrids from Coulomb gauge QCD,''
  Phys.\ Rev.\ D {\bf 78}, 056003 (2008);
  J.~J.~Dudek and E.~Rrapaj,
  %``Charmonium in lattice QCD and the non-relativistic quark-model,''
  Phys.\ Rev.\ D {\bf 78}, 094504 (2008).


\end{thebibliography}
\end{document}